\documentclass[prb,twocolumn,amsmath,amssymb]{revtex4}
\usepackage{amssymb}
\usepackage{amsmath}
\usepackage{graphicx}
\usepackage{float}
\usepackage{ulem}
\usepackage{bm}
\usepackage{dcolumn}
\usepackage{mathrsfs}
\usepackage{subfigure}
\usepackage{braket}
\usepackage{graphicx}
\usepackage{braket}
\usepackage{color}

\begin{document}
\title{Direct prediction of corner state configurations from edge winding numbers in 2D and 3D chiral-symmetric lattice systems}

\author{Linhu Li}
\affiliation{Department of Physics, National University of Singapore, Singapore 117551, Republic of Singapore}
\author{Muhammad Umer}
\affiliation{Department of Physics, National University of Singapore, Singapore 117551, Republic of Singapore}
\author{Jiangbin Gong}
\email{phygj@nus.edu.sg}
\affiliation{Department of Physics, National University of Singapore, Singapore 117551, Republic of Singapore}

\begin{abstract}
Higher-order topological phases feature topologically protected boundary states in lower dimensions. Specifically, the zero-dimensional corner states are protected by the $d$th-order topology of a $d$-dimension system.
In this work, we propose to predict different configurations of corner states from winding numbers defined for one-dimensional edges of the system.
We first demonstrate the winding number characterization with a generalized two-dimensional square lattice belonging to the BDI symmetry class. In addition to the second-order topological insulating phase, the system may also be a nodal point semimetal or a weak topological insulator with topologically protected one-dimensional edge states coexisting with the corner states at zero energy. A three-dimensional cubic lattice with richer configurations of corner states is also studied. We further discuss several experimental implementations of our models with photonic lattices or electric circuits.
\end{abstract}

\maketitle

\section{Introduction}

Over the past decade, topological phases of matter have been one of the most intriguing research topics in condensed matter physics. A $d$-dimensional ($d$D) topological system generally features topological protected states in its $(d-1)$D boundaries, which can be characterized by some topological invariants associated with the bulk. Recently, a new type of topological phases, namely the higher-order topological insulators (HOTI), have attracted growing attention both theoretically \cite{HOTI_science,HOTI_Langbehn,HOTI_Song,HOTI_Benalcazar,HOTI_Schindler,HOTI_Ezawa,HOTI_Kunst,HOTI_Khalaf,HOTI_Lin,HOTI_Xu,HOTI_Xie} and experimentally \cite{HOTI_phononic,HOTI_electronic,HOTI_microwave,HOTI_circuit}. In general, a $n$th-order topological insulator has gapless $(d-n)$D boundary states protected by the system's topology, as well as gapped states in its higher-dimensional boundaries and the bulk. An exception is when $n=d$, where the topologically protected boundary states are some zero-energy corner modes falling between the band gap. 
The remarkable ten-fold way symmetry classification of topological gapped systems \cite{classification_TI1,classification_TI2} has also been extended to describe HOTIs, 
and the topological properties of such systems are determined by its time-reversal symmetry, particle-hole symmetry, and chiral symmetry, and the dimension of the concerned boundaries \cite{HOTI_Khalaf,classification_defect}. On the other hand, spatial symmetries, such as rotation symmetry and mirror symmetry, have also been argued to play a crucial role in determining the existence of topologically protected states at different boundaries.
To topologically characterize the higher-order topological phases, a paradigm for topological invariants based on ``nested" Wilson loops has been introduced \cite{HOTI_science,HOTI_Benalcazar}, while anomalous HOTI phases with a vanishing nested Wilson loop have also been unveiled lately \cite{HOTI_anomalous}. 
More recently, based on an observation of topological transition in a 2D second-order topological insulator in LC circuits, the topological nature of the higher-order boundary states has been further corroborated by direct measurements of the winding numbers of each 1D edge \cite{winding_new}. However, the explicit correspondence between the winding numbers and the corner states is yet to be revealed.

In this work, we propose to predict various configurations of corner states with the winding numbers assigned for 1D edges of our systems, with each corner state corresponding to a different set of winding numbers. In order to illustrate the winding number characterization, we first consider a generalized 2D square lattice belonging to the BDI symmetry class, which may have $0$D topological boundary states (i.e. corner states) described by $Z$-type topological invariants. We find that there are two types of zero energy corner states in this model, which can be systematically characterized by the four winding numbers assigned to each edge of the square lattice. 
On the other hand, these winding numbers do not tell whether the bulk is gapped or not, thus the system may have several different insulating and semimetallic phases. Nevertheless, the configurations of corner states are not directly related to the bulk spectrum, and can be predicted solely from the winding numbers.
Our construction and topological characterization of corner states can also be extended to higher dimensions, and in this paper we study a 3D cubic lattice, which exhibit richer configurations of corner states.

Throughout this paper, we will discuss about boundary states of different dimensions, hence we give a clarification here to avoid possible confusion. We use the word ``boundary" for the general boundaries of a system, which has no specific dimension; and the words ``corner", ``edge", and ``surface" indicate 0D, 1D, and 2D boundaries respectively.
The rest of the paper is organized as follows. In Sec. \ref{sec_2D}, we first introduce the generalized model in the 2D square lattice, and define the winding numbers for each 1D edge of the system. Then we solve the corner states of this system, and study the correspondence between the corner states and the winding numbers. Together with the information of 2D bulk states and 1D edge states of the system, we obtain a phase diagram of the system, showing the possibility of having different types of topological feaures in our system. 
Sec. \ref{sec_3D} introduces a 3D extension of our model, where the winding number characterization also applies.
In Sec. \ref{sec_exp} we propose several physical simulations of our model with different experimental setups of photonic lattices or electrical circuits. Finally, we give a brief summary in Sec. \ref{sec_sum}.

\section{Corner states in 2D square lattices}\label{sec_2D}
\subsection{Hamiltonian and winding numbers}
\begin{figure}
\includegraphics[width=0.8\linewidth]{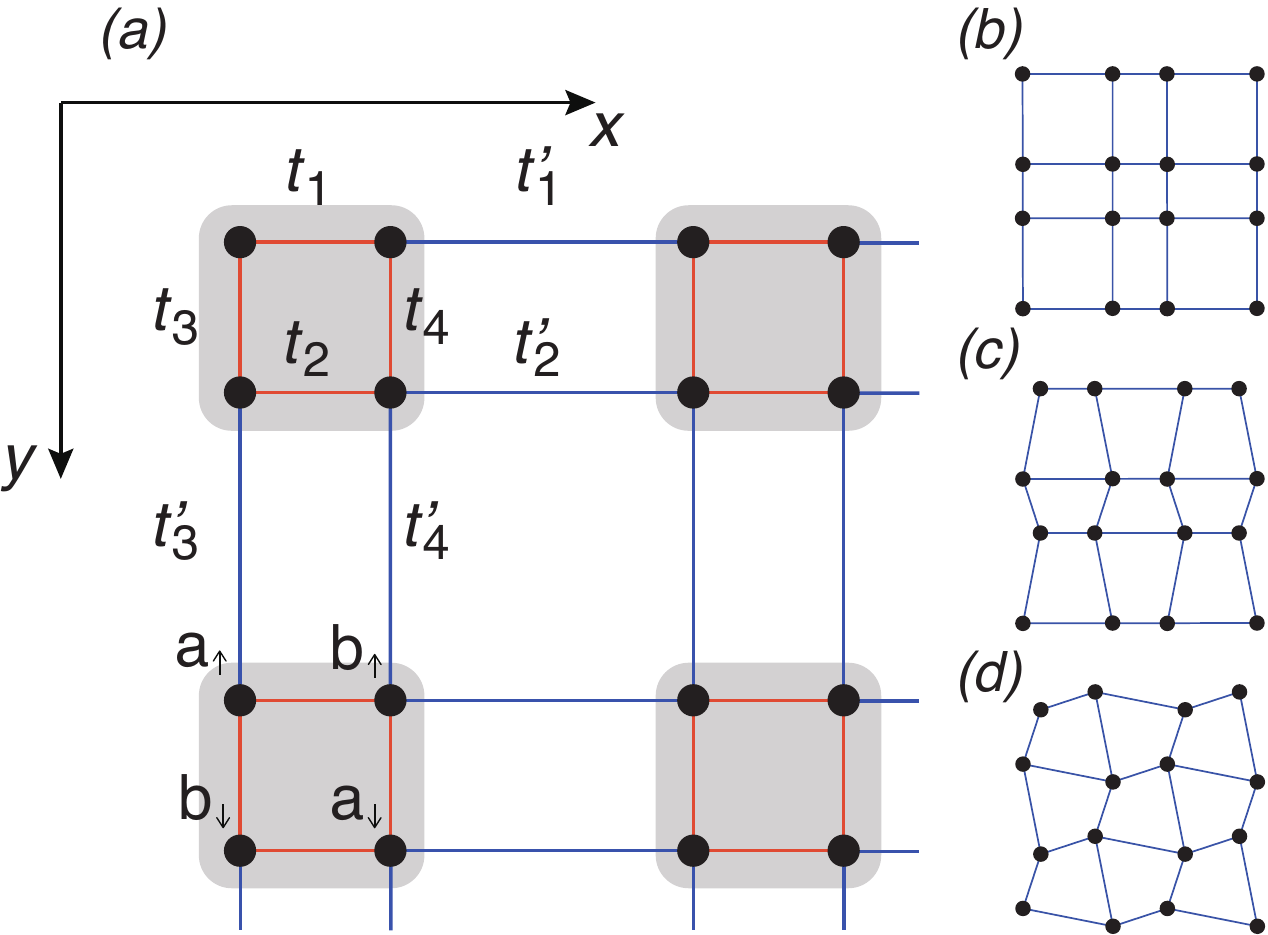}
\caption{Sketches of the lattice of Hamiltonian (\ref{2D_model}). (a) The general lattice structure. (b)-(d) Examples of different situations of the hopping amplitudes. Longer distances between two sites indicate weaker hopping amplitudes. The winding numbers of them are (b) $\nu_1=\nu_2=\nu_3=\nu_4=1$, (c) $\nu_1=0$ and $\nu_2=\nu_3=\nu_4=1$, and (d) $\nu_1=\nu_3=0$ and $\nu_2=\nu_4=1$.}\label{fig1}
\end{figure}

We first consider a square lattice with four sites in each unit cell and only the nearest neighbor hopping, as shown in Fig. \ref{fig1}(a). The sublattices are labeled with two pseudospin-1/2 subspaces $(a,b)$ and $(\uparrow,\downarrow)$.
The Hamiltonian reads
\begin{eqnarray}
H&=&\sum_{x,y}[t_1\hat{a}^{\dagger}_{\uparrow,x,y}\hat{b}_{\uparrow,x,y}+t'_1\hat{b}^{\dagger}_{\uparrow,x,y}\hat{a}_{\uparrow,x+1,y}\nonumber\\
&&+t_2\hat{b}^{\dagger}_{\downarrow,x,y}\hat{a}_{\downarrow,x,y}+t'_2\hat{a}^{\dagger}_{\downarrow,x,y}\hat{b}_{\downarrow,x+1,y}\nonumber\\
&&+t_3\hat{a}^{\dagger}_{\uparrow,x,y}\hat{b}_{\downarrow,x,y}+t'_3\hat{b}^{\dagger}_{\downarrow,x,y}\hat{a}_{\uparrow,x,y+1}\nonumber\\
&&+t_4\hat{b}^{\dagger}_{\uparrow,x,y}\hat{a}_{\downarrow,x,y}+t'_4\hat{a}^{\dagger}_{\downarrow,x,y}\hat{b}_{\uparrow,x,y+1}]+h.c.,\label{2D_model}
\end{eqnarray}
where $\hat{\alpha}_{\beta,x,y}$ denotes the annihilation operator of a fermion on the $\alpha-\beta$ sublattice of the $(x,y)$-th unit cell, with $\alpha=a$ or $b$ and $\beta=$ $\uparrow$ or $\downarrow$. Here we assume no specific spacial symmetry, and the system has eight individual hoppings chosen to be real.

Under the Fourier transformation, the Hamiltonian can be written in momentum space as
\begin{eqnarray}
H_{{\bm k}}= \hat{\psi}^\dagger_{\bm k} \mathcal{H}({\bm k}) \hat{\psi}_{\bm k},
\end{eqnarray}
where $\hat{\psi}_{\bm k}=(\hat{a}_{\uparrow,{\bm k}} ,\hat{a}_{\downarrow,{\bm k}} , \hat{b}_{\uparrow,{\bm k}} ,\hat{b}_{\downarrow,{\bm k}} ,)^T $ and
\begin{eqnarray}
\mathcal{H}({\bm k}) &=& \left(\begin{array}{cccc}
0 & 0 & h_{1} & h_{3} \\
0 & 0 & h^*_{4} & h^*_{2} \\
h^*_{1} & h_{4} & 0 & 0 \\
h^*_{3} & h_{2} & 0 & 0 
\end{array}\right),\label{Hk}
\end{eqnarray}
with $h_{1,2}=t_{1,2}+t'_{1,2}e^{-ik_x}$, and $h_{3,4}=t_{3,4}+t'_{3,4}e^{-ik_y}$. This Hamiltonian is off-diagonal and hence satisfies a chiral symmetry 
\begin{eqnarray}
\tau_3 \mathcal{H}(k) \tau_3=-\mathcal{H}(k), 
\end{eqnarray}
where $\tau_3$ is the third Pauli matrix acting on the $(a,b)$ subspace. 
We note that this model is the most general case with such a chiral symmetry, as it contains four independent complex numbers in the off-diagonal $2\times2$ block,
i.e. eight independent real numbers. 
The Hamiltonian can be expressed in terms of $4\times4$ Gamma matrices defined as $\sigma_i\tau_j$, the direct product of two sets of Pauli matrices. The Hermiticity ensures that each term has a real coefficient, thus the eight real numbers are mapped to the coefficients of all the eight Gamma matrices constructed with either $\tau_1$ or $\tau_2$.
Each of the rest Gamma matrices is constructed either with $\tau_0$ or $\tau_3$, therefore they do not satisfy the chiral symmetry.
Furthermore, this model also satisfies the time-reversal symmetry for spinless system, $\mathcal{H}^*(k)=\mathcal{H}(-k)$, which does not directly protect the topology, but allows us to determine the phase transitions as discussed in Section \ref{sec_phases}.
The system hence belongs to the BDI symmetry class, which does not support topologically nontrivial 1D chiral edge states, but may have robust 0D corner states characterized by one or more $Z$-type topological invariants \cite{HOTI_Khalaf,classification_defect}. Previous studies have shown that the corner states can be topologically characterized by the Wilson loops, which describe the topology of 1D edge states of a 2D SOTI \cite{HOTI_science,HOTI_Benalcazar}. In the lattice system of Fig. \ref{fig1}(a), each 1D edge consists only a pair of alternative hoppings $t_i$ and $t_i'$, thus it can be described by an effective Hamiltonian
\begin{eqnarray}
H_{{\rm 1D},i}=\sum_{n}[t_i\hat{c}^{\dagger}_{n}\hat{d}_{n}+t'_i\hat{d}^{\dagger}_{n}\hat{c}_{n+1}]+h.c.\nonumber\\
\end{eqnarray}
with $n$ denotes $x$ for $i=1,2$ and $y$ for $i=3,4$.
This Hamiltonian describes a 1D Su-Schrieffer-Heeger (SSH) chain \cite{SSH} along either $x$ or $y$ direction.
Due to the translational symmetry regarding unit cells, these 1D SSH chains are also manifested in the bulk of the system, thus a Fourier transform of $H_{{\rm 1D},i}$ gives the same $h_i$ as for the bulk Hamiltonian in Eq. (\ref{Hk}).
To describe the topology of our system, We define four winding numbers for each $h_i$ as
\begin{eqnarray}
\nu_i=\frac{1}{2\pi}\oint \frac{h_{i}^R dh_{i}^I-h_{i}^I dh_{i}^R}{|h_i|^2},
\end{eqnarray}
with $h_{i}^R=\Re[{h_i}]$ and $h_{i}^I=-\Im[{h_i}]$, $i=1,2,3,4$. These winding numbers characterize the 1D topology of the SSH models associated with different 1D edge and different pair of sublattices in our system. For example, $\nu_1$ corresponds to the edge with $y=1$ and the subspace of $(\hat{a}_{\uparrow},\hat{b}_{\uparrow})$. 
In a SSH model, the value of the winding number $\nu_i$ is determined only by the relative strength between the two hoppings $t_i$ and $t'_i$, i.e.  $\nu_i=1$ when $|t_i|<|t'_i|$ and $\nu_i=0$ when $|t_i|>|t'_i|$. By adjusting different pairs of hopping amplitudes, one can obtain various lattice structures with different winding numbers, as illustrated with some examples in Fig. \ref{fig1}(b)-(d). These winding numbers can unambiguously characterize the corner states in our system, as discussed in the following.

 \subsection{Corner state solutions}
To obtain an analytical solution of corner states, we consider the real space eigen-equation for our system, $H\Psi_l=E_l\Psi_l$. Here $E_l$ and $\Psi_l$ are the eigen-energy and eigenfunction $l$-th eigen-state. Assuming 
\begin{eqnarray}
\Psi_l&=&\sum_{x,y}(\psi_{a,\uparrow,x,y}\hat{a}^{\dagger}_{\uparrow,x,y}+\psi_{a,\downarrow,x,y}\hat{a}^{\dagger}_{\downarrow,x,y}\nonumber\\
&&~~~~+\psi_{b,\uparrow,x,y}\hat{b}^{\dagger}_{\uparrow,x,y}+\psi_{b,\downarrow,x,y}\hat{b}^{\dagger}_{\downarrow,x,y})|0\rangle,
\end{eqnarray} 
the eigen-equation leads to
\begin{eqnarray}
&&t_1\psi_{b,\uparrow,x,y}+t'_1\psi_{b,\uparrow,x-1,y}+t_3\psi_{b,\downarrow,x,y}+t'_3\psi_{b,\downarrow,x,y-1}\nonumber\\
&&=E\psi_{a,\uparrow,x,y},\label{EE1}\\
&&t_1\psi_{a,\uparrow,x,y}+t'_1\psi_{a,\uparrow,x+1,y}+t_4\psi_{a,\downarrow,x,y}+t'_4\psi_{a,\downarrow,x,y-1}\nonumber\\
&&=E\psi_{b,\uparrow,x,y},\label{EE2}\\
&&t_2\psi_{a,\downarrow,x,y}+t'_2\psi_{a,\downarrow,x-1,y}+t_3\psi_{a,\uparrow,x,y}+t'_3\psi_{a,\uparrow,x,y+1}\nonumber\\
&&=E\psi_{b,\downarrow,x,y},\label{EE3}\\
&&t_2\psi_{b,\downarrow,x,y}+t'_2\psi_{b,\downarrow,x+1,y}+t_4\psi_{b,\uparrow,x,y}+t'_4\psi_{b,\uparrow,x,y+1}\nonumber\\
&&=E\psi_{a,\downarrow,x,y}.\label{EE4}
\end{eqnarray} 
In order to determine whether there is zero-energy corner states at each corner, we consider the solutions at $E=0$, and first focus on the corner at $x=y=1$ under semi-infinite boundary condition with $x$ and $y$ range from $1$ to infinity. The boundary condition is given by $\psi_{a,\uparrow,x_0,y_0}=\psi_{b,\uparrow,x_0,y_0}=\psi_{a,\downarrow,x_0,y_0}=\psi_{b,\uparrow,x_0,y_0}=0$, with $x_0=0$ or/and $y_0=0$. 
In each of Eqs. (\ref{EE1}) to (\ref{EE4}), the left hand side contains the components of two sublattices, each corresponding to a recurrence relation in a different direction. In order to have corner states, the wave-function must exponentially decay in both $x$ and $y$ directions. 
Substituting the semi-infinte boundary condition to Eqs. (\ref{EE1}) to (\ref{EE4}), we find that there are two types of corner states in the corner of $x=y=1$. The type-1 corner state is given by
\begin{eqnarray}
\psi_{a,\uparrow,x,y}&=&\left(-\frac{t_1}{t'_1}\right)^{x-1}\left(-\frac{t_3}{t'_3}\right)^{y-1},\nonumber\\
\psi_{a,\downarrow,x,y}&=&\psi_{b,\uparrow,x,y}=\psi_{b,\downarrow,x,y}=0,
\end{eqnarray}
which has nonzero amplitudes only on one sublattice, and 
decays exponentially along both $x$ and $y$ directions when $|t_1|<|t'_1|$ and $|t_3|<|t'_3|$, corresponding to the two winding numbers $\nu_1=\nu_3=1$. 
The type-2 corner state is given by
\begin{eqnarray}
\psi_{b,\uparrow,x,y}&=&-t_3\left(-\frac{t'_1}{t_1}\right)^{(x-1)}\left(-\frac{t_4}{t'_4}\right)^{(y-1)},\nonumber\\
\psi_{b,\downarrow,x,y}&=&t_1\left(-\frac{t_2}{t'_2}\right)^{(x-1)}\left(-\frac{t'_3}{t_3}\right)^{(y-1)},\nonumber\\
\psi_{a,\uparrow,x,y}&=&\psi_{a,\downarrow,x,y}=0.
\end{eqnarray}
This solution gives a different corner state with nonzero distribution on two sublattices, under the condition of $|t_i|>|t'_i|$ and $|t_j|<|t'_j|$, with $i=1,3$ and $j=2,4$. 
While the type-1 corner state can be comprehended as a 2D manifestation of the 1D topology of a SSH model, the type-2 corner state is originated in the 2D structure and the coupling between SSH chains along different directions. This can be seen from the recurrence relation at the starting point of the semi-infinite system. In the above example with zero energy, Eq. (\ref{EE1}) reduces to
\begin{eqnarray}
t_1\psi_{b,\uparrow,1,1}+t_3\psi_{b,\downarrow,1,1}=0
\end{eqnarray}
at the corner of $x=y=1$, which indicates either (i) $\psi_{b,\uparrow,1,1}=\psi_{b,\downarrow,1,1}=0$ or (ii) $\psi_{b,\uparrow,1,1}/\psi_{b,\downarrow,1,1}=-t_3/t_1$. The first condition corresponds to the nontrivial topology of two individual SSH chains, and leads to the solution of a type-1 corner state. On the other hand, the second condition has non-vanished wave amplitudes on two sublattices, which are coupled to each other through $t_1$ and $t_3$, two hoppings along $x$ and $y$ directions respectively. This condition lead to the solution of a type-2 corner state, which has no 1D analogue.
While the type-1 corner state is characterized by only two winding numbers, the corner state of type-2 corresponds to all the four winding numbers, i.e. $\nu_1=\nu_3=0$ and $\nu_2=\nu_4=1$, and cannot coexist with the previous one at the same corner.

\begin{figure}
\includegraphics[width=1\linewidth]{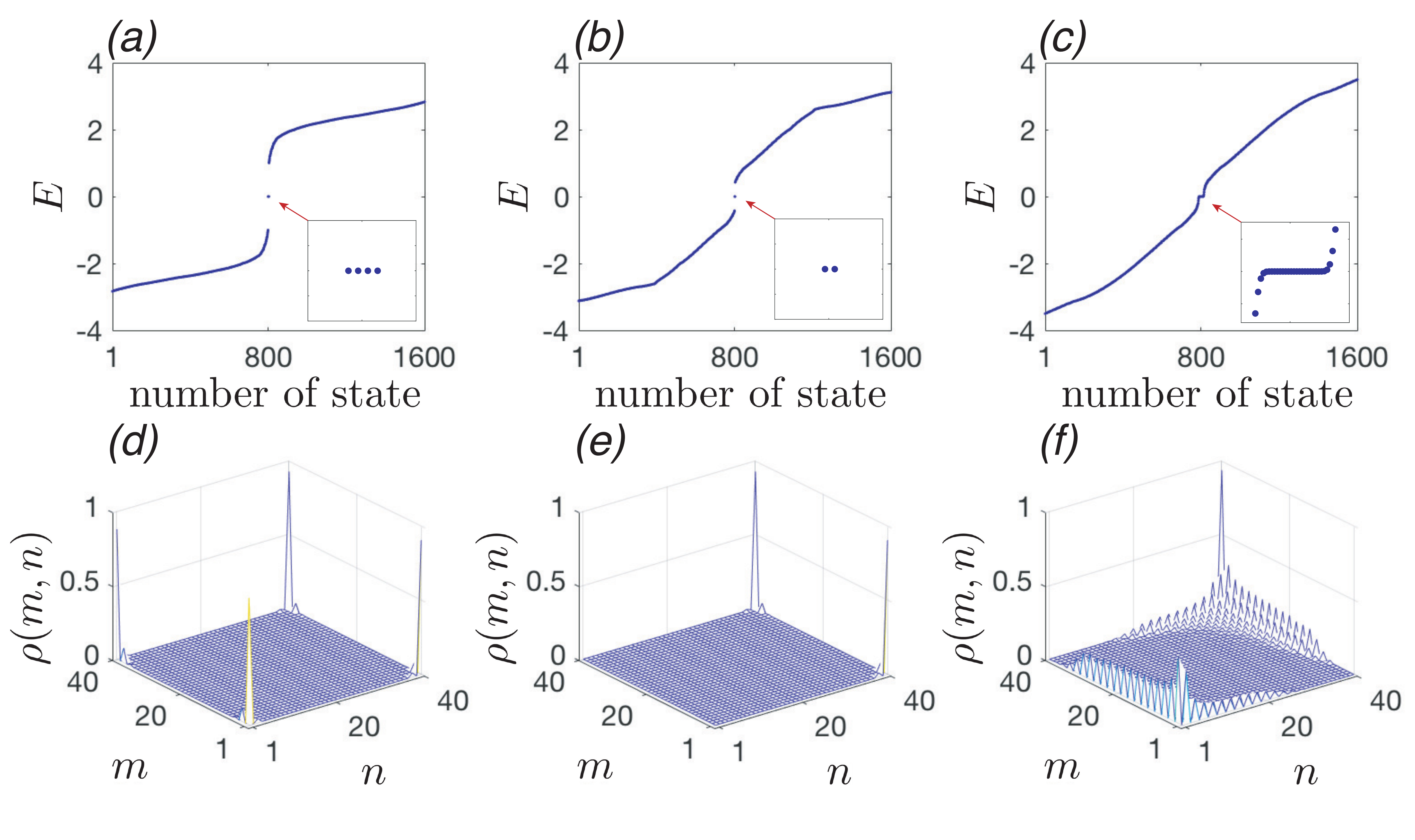}
\caption{Spectra and the collective distribution of zero modes with different parameters for the 2D square lattice, the parameters of which are given by Eqs. \ref{2Dparameter}. The top panels show the spectra with (a) $\delta_1=0.5$, $\delta_2=0.6$, $\delta_3=0.7$ and $\delta_4=0.8$; (b) $\delta_1=-0.5$, $\delta_2=0.6$, $\delta_3=0.7$ and $\delta_4=0.8$; and (c) $\delta_1=-0.5$, $\delta_2=0.6$, $\delta_3=-0.7$ and $\delta_4=0.8$. The collective distribution of zero modes with the same parameters are shown in (d)-(f) respectively.}
\label{fig2}
\end{figure}

In our model, as different corners and hoppings can be exchanged to each other by rotating the lattice, we can obtain the correspondence between different corner states and the winding numbers. As a conclusion, the emergence of type-1 corner states requires the winding numbers of the two edges that intersect at the corresponding corner being one, while the emergence of type-2 corner states requires these two winding numbers being zero, and the rest two being one. A type-1 corner state always localizes at the first lattice site at a corner, and a type-2 corner state localizes at the two sites next to it.
Based on these conditions, in our model the corner states have three different configurations: (i) four type-1 corner states when all the winding numbers equal to one; (ii) two type-1 corner states on two neighbor corners when one winding number is zero and the other three are one; and (iii) a type-1 corner state and a type-2 corner state localized at two diagonal corners respectively, when two winding numbers of a pair of neighbor edges are zero, and the other two are one. 
By visualizing the hopping strength with the distance between two lattices,
these three cases correspond to the sketches in Fig. \ref{fig1}(b)-(d) respectively, and in Fig. \ref{fig2} we illustrate the spectra and distributions of zero modes for systems under each of these conditions. Note that we choose OBC along both $x$ and $y$ directions, therefore the energy bands are given by both the 2D bulk states and 1D edge states, 
and we refer to them as the bulk-edge spectrum hereafter.
We also choose the parameters as
\begin{eqnarray}
t_1=1-\delta_1,&~&t'_1=1+\delta_1;\nonumber\\
t_2=1-\delta_2,&~&t'_2=1+\delta_2;\nonumber\\
t_3=1-\delta_3,&~&t'_3=1+\delta_3;\nonumber\\
t_4=-(1-\delta_4),&~&t'_4=-(1+\delta_4),\label{2Dparameter}
\end{eqnarray}
so that the eight hopping amplitudes are controlled by four parameters $\delta_{1,2,3,4}$. The negative signs of $t_4$ and $t'_4$ are a gauge choice for a $\pi$-flux through each plaquette, which ensures a gap at zero energy in the case with a corner state at each corner \cite{HOTI_science,HOTI_Benalcazar}. The winding numbers are $\nu_i=1$ if $t_i<t'_i$, and $\nu_i=0$ when $t_i>t'_i$. In Fig. \ref{fig2}(d) to (f) we illustrate the collective distribution of all zero modes, which is defined as
\begin{eqnarray}
\rho(m,n)=\sum_{E=0}\psi(m,n)^*\psi(m,n),
\end{eqnarray}
with $\psi(m,n)$ the amplitude of a zero-energy eigen-state at $(m,n)$-th lattice site. Here $m$ and $n$ label the number of lattice sites along $x$ and $y$ direction respectively. 

In Fig. \ref{fig2}(a) and (b) we observe a gapped bulk-edge spectrum with four and two zero-energy corner states as discussed above. In Fig. \ref{fig2}(c) with two corner states localized at diagonal corners, however, the bulk-edge spectrum is gapless. As a matter of fact, we find that the system always has some 1D zero-energy edge states in the case with two diagonal corner states (see Appendix \ref{App_edgestates}). Nevertheless, even in the presence of nonzero distributions along the 1D edges, the distribution in Fig. \ref{fig2}(f) show clear peaks at the two diagonal corners, which distinguish the corner states from the bulk and edge states. 

We also note that the existence of corner states, even in the case of one at each corner, does not require any spacial symmetry of the system, e.g. reflection or rotation. Instead, the $Z$-type topology here is protected solely by the chiral symmetry. However, the localized states at different corners can be related through some ``weak"-spacial symmetries, which only require the system to preserve the same topology after a given spacial transformation, but allow it to be stretched or compressed.
For example, even in the case with four corner states, the system may not be identical to itself after a $C_4$ rotation as each pair of $t_i$ and $t'_i$ can take different values. However, it is topologically equivalent under such a rotation, as all the four winding numbers still equal to one.

\subsection{Phase diagram}\label{sec_phases}
\begin{figure}
\includegraphics[width=1\linewidth]{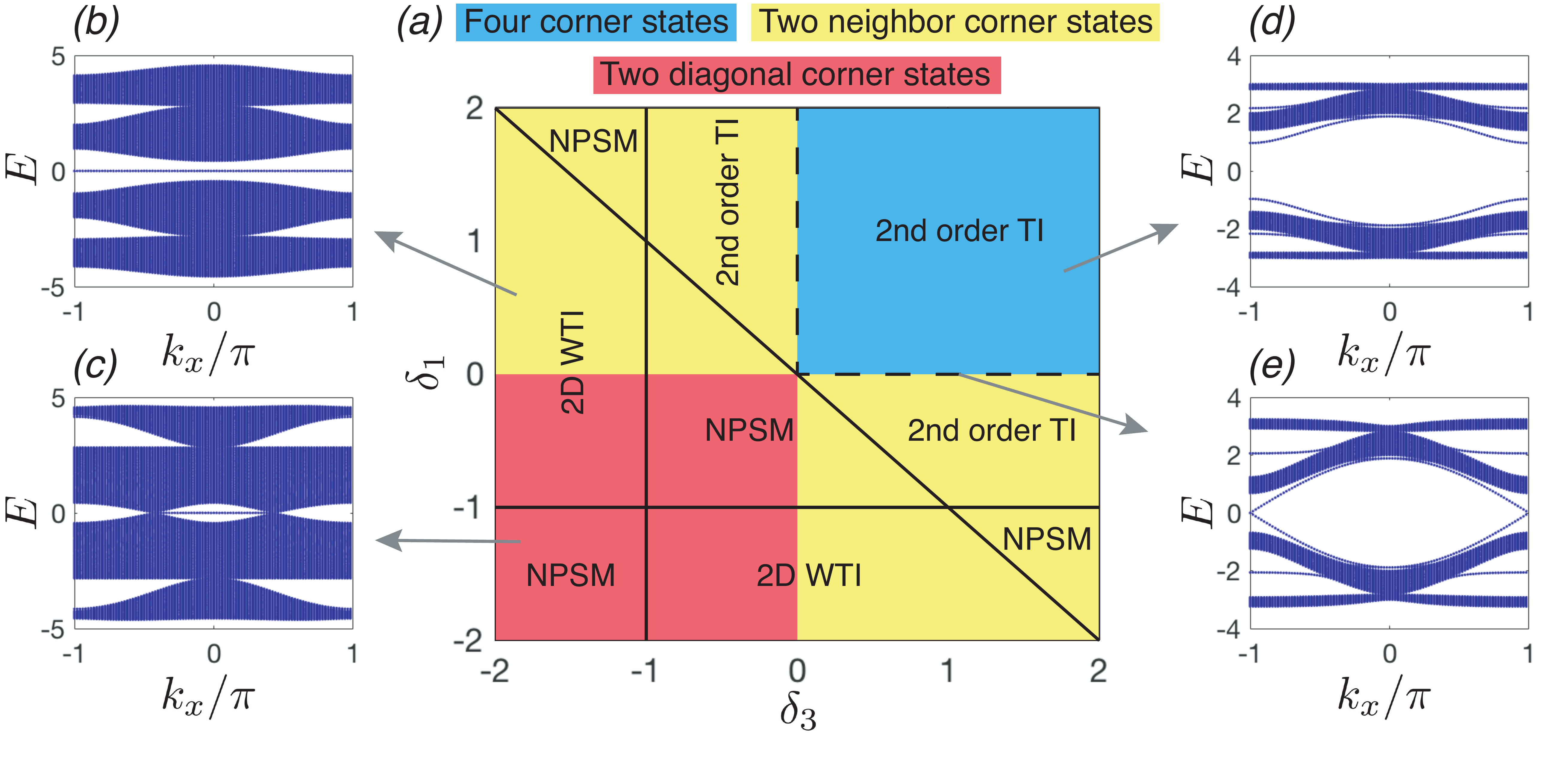}
\caption{A phase diagram of the system with $\delta_2=\delta_4=1$, and the spectra under OBC along $y$ direction in several different parameter regimes. (a) The phase diagram with different phases labeled on it. Different colors show the behavior of corner states, and the solid lines are the boundaries between different phases regarding the bulk and edge states. The dash lines show the boundaries between second-order topological insulators with different types of corner states.
The parameters for the spectra are (b) $\delta_1=0.5$, $\delta_3=-1.5$; (c) $\delta_1=-1.5$, $\delta_3=-1.5$; (d) $\delta_1=0.5$, $\delta_3=0.5$; (e) $\delta_1=0$, $\delta_3=0.5$.}
\label{fig3}
\end{figure}

While the edge winding numbers can unambiguously predict the configurations of corner states, they do not give much information of the bulk-edge spectrum. By taking into account the behaviors of the bulk and edge states, the system will have various types of topological phases.
Here we numerically diagonalize the Hamiltonian, and illustrate a phase diagram with the spectra under OBC along $y$ direction in some typical cases in Fig. \ref{fig3}. As our main concern is the corner states characterized by the winding numbers, we focus on the case with fixed $\delta_2=\delta_4=1$ and choose $\delta_1$ and $\delta_3$ as varing parameters. In this case the system always has a type-1 corner state at $x=N_x$ and $y=N_y$, while $\delta_{1,3}$ determine the situations in the rest three corners. 

In Fig. \ref{fig3}(a), we show the behavior of corner states with different colors, and indicate different phases regarding also the edge and bulk behavior in the diagram.  The three solid lines distinguishing different topological phases are given by
\begin{eqnarray}
\delta_3\delta_4=-1,~\delta_1\delta_2=-1,~{\rm and}~
\delta_1\delta_2+\delta_3\delta_4=0
\end{eqnarray}
respectively, which can be obtained by analyzing the high-symmetric points of the system (see Appendix \ref{App_phaseboundary}).
In Fig. \ref{fig3}(b)-(e) we consider OBC along only $y$ direction, and illustrate some typical examples of the bulk-edge spectrum as a function of $k_x$. Besides the higher order insulating phase, the system can also be an insulator or a nodal point semimetal (NPSM), both with 1D zero-energy edge modes, as shown in Fig. \ref{fig3}(b) and (c) respectively. These zero modes correspond to the 1D topology of the system along $y$ direction with $k_x$ taken as a parameter \cite{footnote}. Specifically, in Fig. \ref{fig3}(b), the system is a 2D insulator with zero Chern number, but hosts topologically protected 1D edge modes at the zero energy. Such a system can be interpreted as a 1D topological insulator layered in $k_x$ direction, and thus forms a 2D analogue of the 3D weak topological insulator (WTI) \cite{FKM_3DTI}.

In Fig. \ref{fig3}(d), we show the bulk-edge spectrum of the system in a second-order topological insulating phase, which show a clear gap between the lower and higher bands (for both edge states and bulk states). In Fig. \ref{fig3}(e), we demonstrate the spectrum at a transition point of two different second-order topological insulating phases. In such case, while the bulk bands remain gapped, the 1D edge states become gapless at $k_x=\pi$, which allows the corner states to emerge or disappear.

\section{Corner states in 3D cubic lattices}\label{sec_3D}
\begin{figure}
\includegraphics[width=1\linewidth]{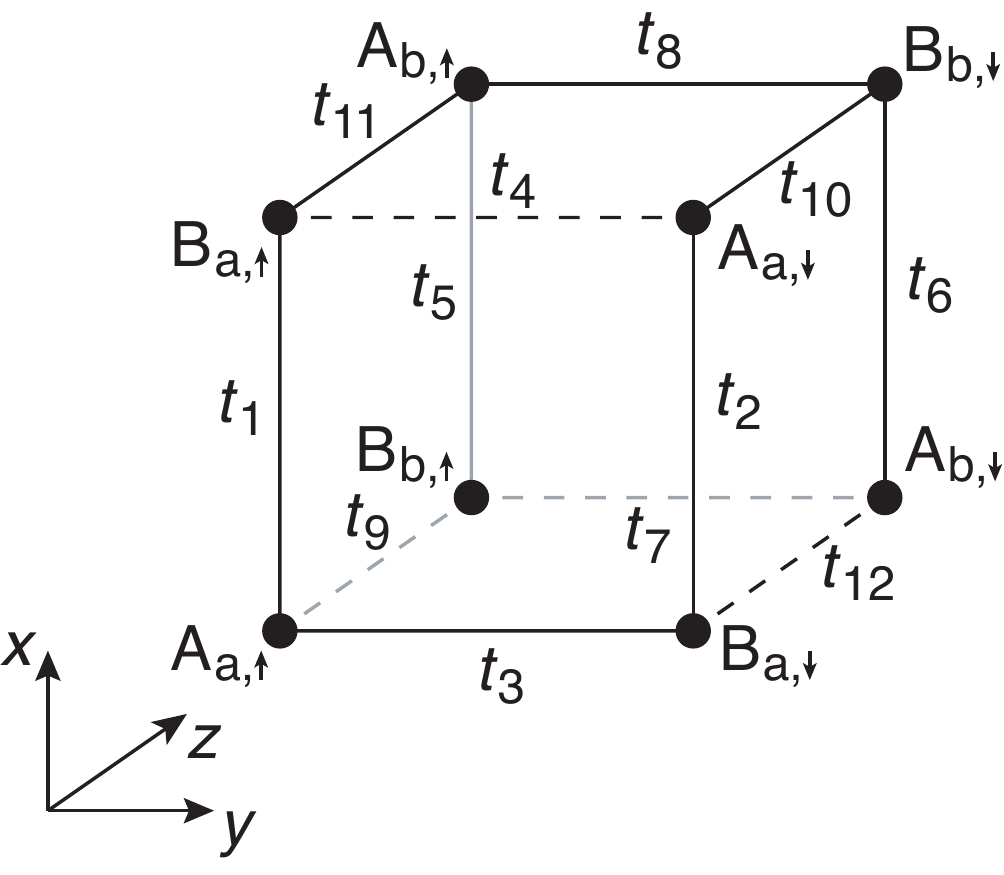}
\caption{A sketch of the 3D lattice. Solid lines are for positive hoppings, dash lines for negative ones, following the choices in Ref. \cite{HOTI_science,HOTI_Benalcazar}. In this sketch we only illustrate a unit cell and the intracell hoppings, while the corresponding intercell hoppings are denoted as $t'_i$ with $i=1,2,...,12$ respectively.}
\label{fig4}
\end{figure}

The construction based on SSH model and the characterization of edge winding numbers can be extended to systems in higher dimensions. A $d$-dimensional cubic lattice has $2^d$ corners and $d 2^{d-1}$ edges, while each corner is the crossing points of $d$ edges and can be described by $d$ winding numbers. Therefore the configurations of corner states are much richer in higher dimensions, and here we take a 3D cubic lattice as an example. We assume that the hoppings take alternative values along each direction, hence a 3D cubic lattice has eight sublattice in a unit cell, as illustrated in Fig. \ref{fig4}. Without any further restriction, there are twelve pairs of hoppings, denoted as $t_i$ and $t'_i$, with $i=1,2,...,12$. In Fig. \ref{fig4} we only show a single unit cell and the intracell hoppings of $t_i$, while each $t'_i$ connects the same sublattices, but between two neighboring unit cells.
The real space Hamiltonian of the 3D cubic lattice reads:
\begin{eqnarray}
H&=&\sum_{{\bm n}}[t_1\hat{A}^{\dagger}_{a,\uparrow,{\bm n}}\hat{B}_{a,\uparrow,{\bm n}}+t'_1\hat{B}^{\dagger}_{a,\uparrow,{\bm n}}\hat{A}_{a,\uparrow,{\bm n}+\hat{x}}\nonumber\\
&&+t_2\hat{B}^{\dagger}_{a,\downarrow,{\bm n}}\hat{A}_{a,\downarrow,{\bm n}}+t'_2\hat{A}^{\dagger}_{a,\downarrow,{\bm n}}\hat{B}_{a,\downarrow,{\bm n}+\hat{x}}\nonumber\\
&&+t_3\hat{A}^{\dagger}_{a,\uparrow,{\bm n}}\hat{B}_{a,\downarrow,{\bm n}}+t'_3\hat{B}^{\dagger}_{a,\downarrow,{\bm n}}\hat{A}_{a,\uparrow,{\bm n}+\hat{y}}\nonumber\\
&&+t_4\hat{B}^{\dagger}_{a,\uparrow,{\bm n}}\hat{A}_{a,\downarrow,{\bm n}}+t'_4\hat{A}^{\dagger}_{a,\downarrow,{\bm n}}\hat{B}_{a,\uparrow,{\bm n}+\hat{y}}\nonumber\\
&&+t_5\hat{B}^{\dagger}_{b,\uparrow,{\bm n}}\hat{A}_{b,\uparrow,{\bm n}}+t'_5\hat{A}^{\dagger}_{b,\uparrow,{\bm n}}\hat{B}_{b,\uparrow,{\bm n}+\hat{x}}\nonumber\\
&&+t_6\hat{A}^{\dagger}_{b,\downarrow,{\bm n}}\hat{B}_{b,\downarrow,{\bm n}}+t'_6\hat{B}^{\dagger}_{b,\downarrow,{\bm n}}\hat{A}_{b,\downarrow,{\bm n}+\hat{x}}\nonumber\\
&&+t_7\hat{B}^{\dagger}_{b,\uparrow,{\bm n}}\hat{A}_{b,\downarrow,{\bm n}}+t'_7\hat{A}^{\dagger}_{b,\downarrow,{\bm n}}\hat{B}_{b,\uparrow,{\bm n}+\hat{y}}\nonumber\\
&&+t_8\hat{A}^{\dagger}_{b,\uparrow,{\bm n}}\hat{B}_{b,\downarrow,{\bm n}}+t'_8\hat{B}^{\dagger}_{b,\downarrow,{\bm n}}\hat{A}_{b,\uparrow,{\bm n}+\hat{y}}\nonumber\\
&&+t_9\hat{A}^{\dagger}_{a,\uparrow,{\bm n}}\hat{B}_{b,\uparrow,{\bm n}}+t'_9\hat{B}^{\dagger}_{b,\uparrow,{\bm n}}\hat{A}_{a,\uparrow,{\bm n}+\hat{z}}\nonumber\\
&&+t_{10}\hat{A}^{\dagger}_{a,\downarrow,{\bm n}}\hat{B}_{b,\downarrow,{\bm n}}+t'_{10}\hat{B}^{\dagger}_{b,\downarrow,{\bm n}}\hat{A}_{a,\downarrow,{\bm n}+\hat{z}}\nonumber\\
&&+t_{11}\hat{B}^{\dagger}_{a,\uparrow,{\bm n}}\hat{A}_{b,\uparrow,{\bm n}}+t'_{11}\hat{A}^{\dagger}_{b,\uparrow,{\bm n}}\hat{B}_{a,\uparrow,{\bm n}+\hat{z}}\nonumber\\
&&+t_{12}\hat{B}^{\dagger}_{a,\downarrow,{\bm n}}\hat{A}_{b,\downarrow,{\bm n}}+t'_{12}\hat{A}^{\dagger}_{b,\downarrow,{\bm n}}\hat{B}_{a,\downarrow,{\bm n}+\hat{z}}]+h.c.,\nonumber\\
\label{model_3D}
\end{eqnarray}
here ${\bm n}=(x,y,z)$ denotes the ${\bm n}$th unit cell, and ${\bm n}+\hat{\alpha}$ denotes the unit cell next to it along $\alpha$ direction.

Under a Fourier transform, the Hamiltonian can be written in momentum space as
\begin{eqnarray}
H_{{\bm k}}= \hat{\psi}^\dagger_{\bm k} \mathcal{H}({\bm k}) \hat{\psi}_{\bm k},
\end{eqnarray}
where 
\begin{eqnarray}
\hat{\psi}_{\bm k}&=&(\hat{A}_{a,\uparrow,{\bm k}} ,\hat{A}_{a,\downarrow,{\bm k}} , \hat{A}_{b,\uparrow,{\bm k}} ,\hat{A}_{b,\downarrow,{\bm k}},\nonumber\\
&&~~~~\hat{B}_{a,\uparrow,{\bm k}} ,\hat{B}_{a,\downarrow,{\bm k}} , \hat{B}_{b,\uparrow,{\bm k}} ,\hat{B}_{b,\downarrow,{\bm k}})^T
\end{eqnarray}
and
\begin{eqnarray}
\mathcal{H}({\bm k}) &=& \left(\begin{array}{cc}
0 & \mathcal{M}({\bm k})  \\
\mathcal{M}^{\dagger}({\bm k})  & 0 
\end{array}\right),\\
\mathcal{M}({\bm k}) &=& \left(\begin{array}{cccc}
h_{1} & h_{3} & h_{9} & 0 \\
h^*_{4} & h^*_{2} & 0 & h_{10}\\
h^*_{11} & 0 & h^*_{5} & h_{8}\\
0 & h^*_{12} & h^*_{7} & h_{6}
\end{array}\right),\nonumber
\end{eqnarray}
with $h_i=t_i+t'_i e^{-i k_{\alpha}}$, $\alpha=x$, $y$, or $z$, determined by the hopping direction of $h_i$. Following the method discussed for the 2D case, we now define $12$ edge winding numbers $\nu_i$ for each $h_i$, and each winding number describes a SSH-like edge of the 3D lattice. 
This model also satisfies the time-reversal symmetry $\mathcal{H}^*({\bm k}) =\mathcal{H}(-{\bm k})$, and the chiral symmetry $\tau_3 \mathcal{H}({\bm k}) \tau_3=-\mathcal{H}({\bm k})$ as it is block-diagonal. Here $\tau_3$ is the third Pauli matrix acting on the subspace of $(A,B)$. Nevertheless, we shall note that in contrast to the 2D system we study, this 3D cubic lattice is not the most general case with the chiral symmetry, as the vanishing off-diagonal elements in $\mathcal{M}({\bm k})$ allow us to add extra terms without breaking the symmetry. These elements correspond to some long-range couplings between diagonal sublattices (e.g. $\hat{A}^{\dagger}_{a,\uparrow,{\bm n}}\hat{B}_{b,\downarrow,{\bm n}}$) in the 3D cubic lattice.

As in the 2D case, there are more than one type of zero-energy corner states localized at different numbers of lattice sites within a unit cell, which can also be obtained by taking semi-infinite boundary condition of the 3D lattice. For instance, at the corner with $(x,y,z)=(1,1,1)$, a type-1 corner state localized at the first lattice site of this corner is given by
\begin{eqnarray}
\psi_{A,a,\uparrow,x,y,z}=\left(-\frac{t_1}{t'_1}\right)^{x-1}\left(-\frac{t_3}{t'_3}\right)^{y-1}\left(-\frac{t_9}{t'_9}\right)^{z-1}
\end{eqnarray}
with vanishing wave amplitudes on other sublattices, providing that $\nu_1=\nu_3=\nu_9=1$.
Similarly, a type-2 corner state localized at the three neighboring sites of the first lattice site can be expressed as
\begin{eqnarray}
\psi_{B,a,\uparrow,x,y,z}=-t_9\left(-\frac{t'_1}{t_1}\right)^{x-1}\left(-\frac{t_4}{t'_4}\right)^{y-1}\left(-\frac{t_{11}}{t'_{11}}\right)^{z-1};\nonumber\\
\psi_{B,a,\downarrow,x,y,z}=-t_9\left(-\frac{t_2}{t'_2}\right)^{x-1}\left(-\frac{t'_3}{t_3}\right)^{y-1}\left(-\frac{t_{12}}{t'_{12}}\right)^{z-1};\nonumber\\
\psi_{B,b,\uparrow,x,y,z}=(t_1+t_3)\left(-\frac{t_5}{t'_5}\right)^{x-1}\left(-\frac{t_7}{t'_7}\right)^{y-1}\left(-\frac{t'_9}{t_9}\right)^{z-1}\nonumber\\
\end{eqnarray}
with vanishing wave amplitudes on other sublattices, and the edge winding numbers shall satisfy $\nu_1=\nu_3=\nu_9=0$ and $\nu_2=\nu_4=\nu_5=\nu_7=\nu_{11}=\nu_{12}=1$.

\begin{figure*}
\includegraphics[width=1\linewidth]{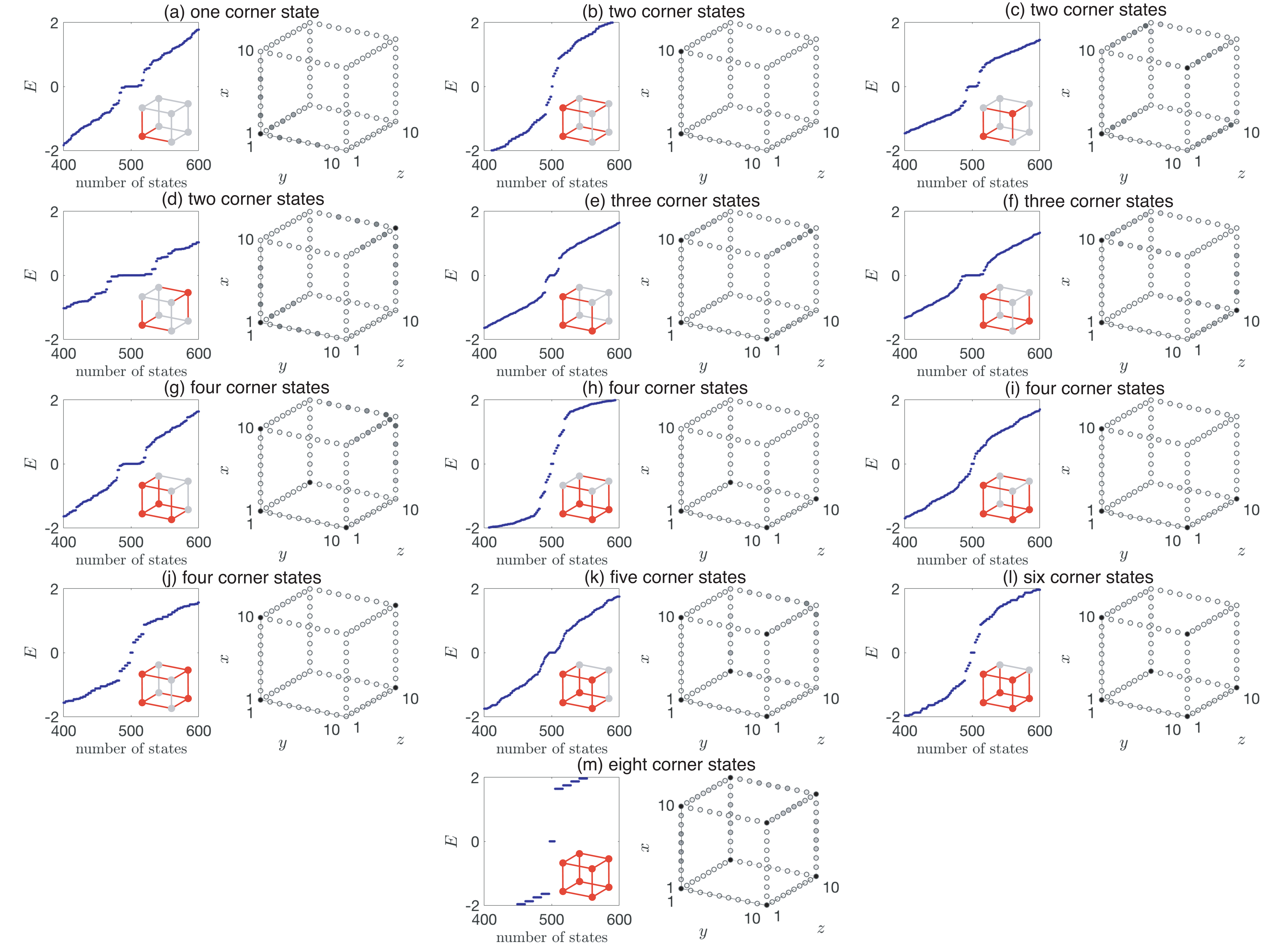}
\caption{Different configurations of the corner states in the 3D cubic lattice. The numbers of the first type of corner states are indicated in the figures. The insert of each panel shows a sketch of the configuration of corner states, with red circles and lines indicate the corners with localized states and the edges with winding numbers $\nu_i=1$, and gray ones for the rest.
The left figure in each panel shows the central part of the spectrum under OBC, which has a gap near zero energy and some in-gap corner states in panels (b), (h), (i) (j), (l), and (m);
the right one shows the collective distribution of zero modes ($E<10^{-4}$ in numerical calculations) along the edges, 
the amplitude of which is indicated by the shade, the darker ones are for larger amplitude. The bulk distributions are omitted for a clearer view. The parameters are chosen as $\delta_i=0.8$ for $\nu_i=1$, and $\delta_i=-0.8$ for $\nu_i=0$. The system's size are chosen as $5\times5\times5$ unit cells.
}
\label{fig5}
\end{figure*}

Compared with the 2D case discussed previously, the 3D structure provides more complicated combinations of couplings between different SSH-like edges, and may hosts even richer types of corner states. For example, instead of being localized at all the three neighboring sites of the first lattice site of a corner, a corner state can also be localized at only two of the three sites, corresponding to different values of the edge winding numbers. However, to list all the possibilities is rather tedious and provides no further insight of the correspondence between corner states and edge winding numbers.
For simplicity, here we focus only on the type-1 corner states, which localized around the exact corners of the 3D lattices. 
Compared with the 2D case, the eight corners and twelve winding numbers of the 3D cubic lattice provide more possible configurations of the corner states, and we demonstrate the thirteen different configurations of corner states in Fig. \ref{fig5}. Any other possibility can be obtained from these configurations with some rotations and/or reflections of the 3D lattice.
Here we choose the parameters as $t_i=\eta(1-\delta_i)$ and $t_i=\eta(1+\delta_i)$, with $\eta=-1$ for $i=4$, $7$, $9$, or $12$, and $\eta=1$ for the rest. The negative signs give a $\pi$-flux through each facet of the cubic lattice, which open a gap in the case with eight corner states, one at each corner respectively \cite{HOTI_science,HOTI_Benalcazar}.

Under these choices, the winding numbers are $\nu_i=1$ for $\delta_i>0$, and $\nu_i=0$ for $\delta_i<0$, with $i=1,2,...,12$. As each type-1 corner state corresponds to three winding numbers, 
for each configuration, there may be more than one choice of the nonzero winding numbers. For example, in order to have a single type-1 corner state at $x=y=z=1$, the edge winding numbers must satisfy $\nu_1=\nu_3=\nu_9=1$. If all the other winding numbers are zero, each of the rest corners has at most one neighboring edge with a nonzero winding number, hence
the system can also have at least one other winding number being nonzero without generating more type-1 corner states.
These different choices may result in different bulk-surface-edge spectra, and we choose the one where the spectrum is gapped at zero energy for most cases. In the rest where we do not find a gap at zero energy, the collective distribution of zero modes also show clear localizations at the corners with three corresponding winding numbers being $1$. 
Note that in some cases there are also other types of corner states, which do not localize at the exact corner of the lattice.

\section{experimental realization}\label{sec_exp}
The key element to realize different types of corner states in our model, is to have separately tunable hoppings, so that each winding number can be adjusted individually. In this regard, we discuss about several possible realization of our model with different experimental setups of photonic lattice and electrical circuit system, as listed below.

\subsection{Waveguide arrays}
\begin{figure}
\includegraphics[width=1\linewidth]{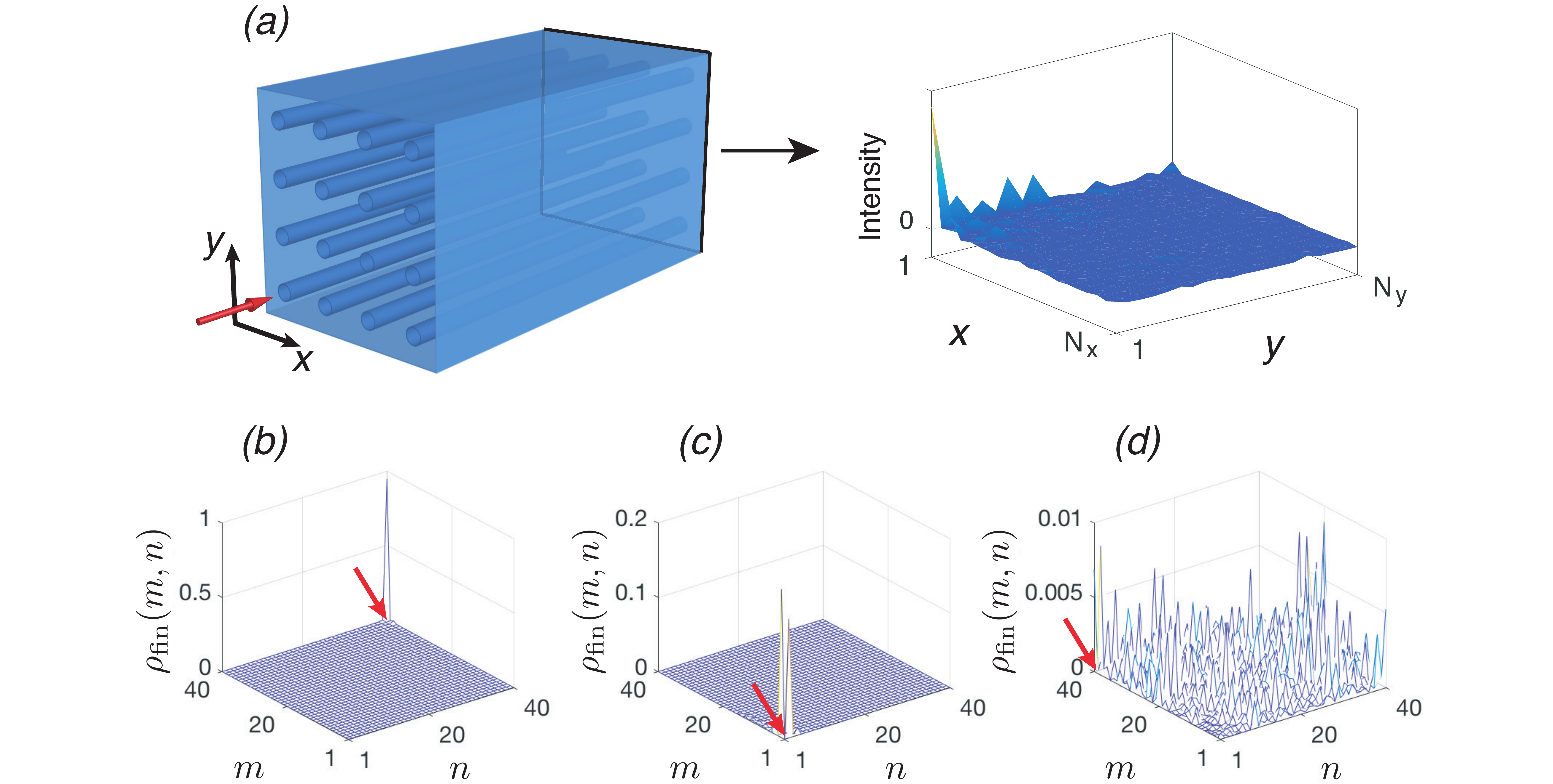}
\caption{Observation of corner states in a 2D photonic lattice. (a) A sketch of the arranged waveguides. The red arrow indicates the injecting light, and the right panel show an example of the outgoing intensity distribution. (b)-(d) the distribution $\rho_{\rm fin}\rho_{\rm fin}(m,n)$ of the final state, with initial states chosen as soliton states in different corners, indicated by the red arrows. The parameters are $\delta_1=-0.5$, $\delta_2=0.6$, $\delta_3=-0.7$ and $\delta_4=0.8$, with the evolving time $T=100$.}
\label{fig6}
\end{figure}

One promising realization of the 2D model is a photonic lattice composed by a series of single-mode waveguides [as shown in Fig. \ref{fig6}(a)], where the hopping amplitudes are determined by the distance between two neighbor waveguides \cite{waveguide1,waveguide2,waveguide3,waveguide4}. The negative sign of the hopping amplitudes can be obtained by inserting an extra waveguide in the center of two original waveguides, and adjusting their onsite potentials (proportional to the refraction index of the waveguides) \cite{HOTI_science,waveguide5}.
Therefore the different phases in our system can be obtained by arranging the waveguides in different configurations as in Fig. \ref{fig1}(b)-(d) (without showing the extra waveguides which induce negative hopping amplitudes). The corner states can then be detected by injecting light into a single waveguide and measuring the outgoing intensity distribution.

To numerically simulate the propagation of the injected light, we consider an initial state $\Psi_{\rm ini}$ representing the injecting light, and its time-evolution in our system. The final state $\Psi_{\rm fin}$ at time $T$ is given by
\begin{eqnarray}
\Psi_{{\rm fin}}=e^{-i H T}\Psi_{{\rm ini}},
\end{eqnarray}
and its distribution
\begin{eqnarray}
\rho_{\rm fin}(m,n)=\psi_{\rm fin}(m,n)^*\psi_{\rm fin}(m,n)
\end{eqnarray}
gives the outgoing intensity of the system. In Fig. \ref{fig6}(b)-(d) we illustrate some examples of the distribution of final states in the case with two diagonal corner states as in Fig. \ref{fig2}(c), where it is seen to have both types of corner states in two diagonal corners, and no corner states in the rest two. 
The initial states are chosen to have nonzero amplitudes only in the unit cells in different corners, as indicated by the red arrows in the figures. 
In order to observe the corner modes, we also choose the initial states to have an appreciable overlapping with the corresponding corner modes.
In \ref{fig6}(b) and (c) the final states are seen to localize at different corners with different distributions, corresponding to the type-1 and type-2 corner states respectively. Under the same parameters, there is no corner state at the rest two corners, so that light injecting these corners shall result in a extended outgoing intensity distribution, as in Fig. \ref{fig6}(d). Note that in this example the system is a nodal point semimetal in the bulk, and has a large number of gapless 1D edge states. Nevertheless, our resutls show that the corner states are robust through time evolution, and can be clearly detected.

Note however, the method here cannot be directly extanded to simulate a 3D lattice, as the light already takes one spatial dimension to propagate.

\subsection{Coupled optical ring resonators}
An alternative route to realize a 2D or 3D lattice system with photons is to use coupled optical ring resonators \cite{CORR1,CORR2,CORR3}. In such simulation, each optical ring resonator represents a lattice site, and these ``site-resonators" are coupled by another set of ring resonators, namely the ``link-resonators". The hopping amplitude is given by the coupling rate between two site-resonators, and can be controlled by tuning the width and the coupling gap of the waveguides composing the link-resonators. Furthermore, by spatially shifting the link-resonators along one axis, say $x$, the photons hopping along $y$ axis can acquire a phase $x\phi$, realizing a synthetic magnetic field with a flux $\phi$. The negative hopping amplitudes in Eqs. (\ref{2Dparameter}) can be obtained by choosing $\phi=\pi$, allowing the system to have gapped bulk-edge spectrum and corner states in the gap as in Fig. \ref{fig2}(a) and (b). The eigen-energies of different states correspond to the frequency of an incoming photon, and the corner states can be detected by launching a light at a certain frequency and measuring the relative amount of light scattered from each site, which effectively gives the distribution of states. However, it may be difficult to identify the type-2 corner states through this measurement, as they always co-exist with some 1D edge states with the same energy, i.e. the same frequency.




\subsection{Electrical circuits}
Besides the photonic lattices, our model can also be simulated with electrical circuit lattices composed by inductors and capacitors, whose circuit Laplacian is analogous to the Hamiltonian of a tight-binding model \cite{Circuit1,Circuit2,Circuit3,HOTI_circuit}. In a circuit lattice, each lattice site is represented by a node of the circuit, and different nodes are connected either by inductors or capacitors. The hopping amplitude in a tight-binding Hamiltonian is identified by the capacitance between two nodes, and an inductor effectively gives a negative capacitance. Therefore the different signs of hopping amplitudes in Eqs. (\ref{2Dparameter})  can also be realized in an electrical circuit lattice. In such a scenario, the corner states can be detected by the topological boundary resonances in the corner impedance of the circuit. As a matter of fact, the 2D square lattice of Eq. (\ref{2D_model}) with four corner states [i.e. the lattice structure of Fig. \ref{fig1}(b)] has already been realized in such systems \cite{HOTI_circuit}. The other lattice structures of Fig. \ref{fig1}(c) and (d) can be obtained by simply using different inductors and capacitors to connect the circuit network, and one can achieve a clearly resolvable corner state resonance on the superimposed resistive background of the bulk states, which indicates that the corner states can be detected in this way even when the bulk-edge states are gapless, as in Fig. \ref{fig2}(c). Furthermore, 
the circuit system has the advantage that in principle it can simulate lattice model in arbitrary dimension, which provides a route to realize the vast variety of configurations of corner states in higher dimensions.

\section{summary}\label{sec_sum}
In this work we propose a direct prediction of different configurations of corner states from winding numbers defined for the 1D edges of a system.
We first study a generalized 2D square lattice with SSH-like edges.
This model has both the chiral symmetry and the time-reversal symmetry, and hosts two types of corner states, each corresponding to a different combination of the edge winding numbers. 
We also unveil that while the type-1 corner states can be comprehended as a 2D manifestation of the 1D topology of a SSH model, the type-2 corner states are rooted in the 2D structure and the coupling between different SSH-like edges along different directions.
On the other hand, the behavior of 2D bulk and 1D edge states of the system is not directly related to the edge winding numbers. 
The system may be a 2D weak topological insulator or a nodal point semimetal in some parameter regimes, but the transition between these phases does not necessarily change the winding numbers and the configurations of corner states. 
Next, we study a 3D cubic lattice of the same symmetry class, which also hosts different types of corner states. In order to illustrate the correspondence between edge winding numbers and corner states in a higher dimension, we systematically list all the thirteen possible configurations of the type-1 corner states in this model.
Finally, we propose a scheme to realize the 2D model with waveguide arrays, and also discuss some other possible experimental realizations of coupled optical ring resonators and electrical circuits, with electrical circuits being most promising in simulating 3D and even higher-dimensional systems.






\acknowledgments
J. G. acknowledges support from Singapore Ministry of Education Academic Research Fund Tier I (WBS No. R- 144-000-353-112) and by Singapore NRF Grant No. NRF- NRFI2017-04 (WBS No. R-144-000-378-281). 

\bigskip

\onecolumngrid
\appendix
\section{edge states in the case with two diagonal corner states}\label{App_edgestates}

In this appendix we show that the system always has some 1D edge states in the case with two diagonal corner states in our 2D model. First we take $k_x$ as a parameter and consider a effective 1D system along $y$ direction. Thus the Hamiltonian satisfies a effective chiral symmetry, $\tau_3h({k_y})\tau_3=-h({k_y})$, and no time reversal symmetry or particle-hole symmetry. Therefore the effective 1D system belongs to the AIII class, which may have zero mode edge states. 

In order to solve the 1D edge states, we take OBC along $y$ direction, and the Hamiltonian reads
\begin{eqnarray}
H_{k_x-y}&=&\sum_{k_x,y}[(t_1+t'_1e^{-ik_x})\hat{a}^{\dagger}_{\uparrow,k_x,y}\hat{b}_{\uparrow,k_x,y}
+(t_2+t'_2e^{ik_x})\hat{a}^{\dagger}_{\downarrow,k_,x,y}\hat{b}_{\downarrow,k_x,y}]+h.c.\nonumber\\
&&+\sum_{k_x,y}[t_3\hat{a}^{\dagger}_{\uparrow,k_x,y}\hat{b}_{\downarrow,k_x,y}+t'_3\hat{b}^{\dagger}_{\downarrow,k_x,y}\hat{a}_{\uparrow,k_x,y+1}
+t_4\hat{b}^{\dagger}_{\uparrow,k_x,y}\hat{a}_{\downarrow,k_x,y}+t'_4\hat{a}^{\dagger}_{\downarrow,k_x,y}\hat{b}_{\uparrow,k_x,y+1}]+h.c..\label{kxy}
\end{eqnarray}
Assuming the wave-functions under OBC along $y$ direction as follows;
\begin{eqnarray}
\Psi_{y,m}=\sum_{y}(\psi_{a,\uparrow,k_x,y}\hat{a}^{\dagger}_{\uparrow,k_x,y}+\psi_{a,\downarrow,k_x,y}\hat{a}^{\dagger}_{\downarrow,k_x,y}+\psi_{b,\uparrow,k_x,y}\hat{b}^{\dagger}_{\uparrow,k_x,y}+\psi_{b,\downarrow,k_x,y}\hat{b}^{\dagger}_{\downarrow,k_x,y})|0\rangle,
\end{eqnarray} 
the eigen equation $H_y(k_x)\Psi_{y,m}(k_x)=E_{y,m}(k_x)$ leads to
\begin{eqnarray}
(t_1+t'_1e^{-ik_x})\psi_{b,\uparrow,k_x,y}+t_3\psi_{b,\downarrow,k_x,y}+t'_3\psi_{b,\downarrow,k_x,y-1}=E\psi_{a,\uparrow,k_x,y},\label{EE1_kxy}\\
(t_1+t'_1e^{ik_x})\psi_{a,\uparrow,k_x,y}+t_4\psi_{a,\downarrow,k_x,y}+t'_4\psi_{a,\downarrow,k_x,y-1}=E\psi_{b,\uparrow,k_x,y},\label{EE2_kxy}\\
(t_2+t'_2e^{-ik_x})\psi_{a,\downarrow,k_x,y}+t_3\psi_{a,\uparrow,k_x,y}+t'_3\psi_{a,\uparrow,k_x,y+1}=E\psi_{b,\downarrow,k_x,y},\label{EE3_kxy}\\
(t_2+t'_2e^{ik_x})\psi_{b,\downarrow,k_x,y}+t_4\psi_{b,\uparrow,k_x,y}+t'_4\psi_{b,\uparrow,k_x,y+1}=E\psi_{a,\downarrow,k_x,y}.\label{EE4_kxy}
\end{eqnarray} 
Here we consider only the zero-energy states, i.e. $E=0$. By substituting these equations to each other, we can obtain the transfer matrices for each sublattices:
\begin{eqnarray}
F_{k_x,b}\left(\begin{array}{c}
\psi_{b,y}\\
\psi_{b,y-1}
\end{array}\right)=\left(\begin{array}{cc}
\frac{(t_1+t'_1e^{-ik_x})(t_2+t'_2e^{ik_x})-t_3t_4-t'_3t'_4
}{t_3t'_4}
 & \frac{-t'_3t_4}{t_3t'_4} \\
1 & 0 
\end{array}\right)\left(\begin{array}{c}
\psi_{b,y}\\
\psi_{b,y-1}
\end{array}\right)=\left(\begin{array}{c}
\psi_{b,y+1}\\
\psi_{b,y}
\end{array}\right),\nonumber\\
F_{k_x,a}\left(\begin{array}{c}
\psi_{a,y}\\
\psi_{a,y-1}
\end{array}\right)=\left(\begin{array}{cc}
\frac{(t_1+t'_1e^{ik_x})(t_2+t'_2e^{-ik_x})-t_3t_4-t'_3t'_4
}{t'_3t_4}
 & \frac{-t_3t'_4}{t'_3t_4} \\
1 & 0 
\end{array}\right)\left(\begin{array}{c}
\psi_{a,y}\\
\psi_{a,y-1}
\end{array}\right)=\left(\begin{array}{c}
\psi_{a,y+1}\\
\psi_{a,y}
\end{array}\right).\nonumber\\
\end{eqnarray} 
Here we neglect the notation of pseudo-spin ($\uparrow$ and $\downarrow$), as the transfer matrices are the same for them. Similarly, we can also take $k_y$ as a parameter and write down the transfer matrices for each sublattices along $x$ direction,
\begin{eqnarray}
F_{k_y,b}\left(\begin{array}{c}
\psi_{b,x}\\
\psi_{b,x-1}
\end{array}\right)=\left(\begin{array}{cc}
\frac{(t_3+t'_3e^{-ik_xy})(t_4+t'_4e^{ik_y})-t_1t_2-t'_1t'_2
}{t_1t'_2}
 & \frac{-t'_1t_2}{t_1t'_2} \\
1 & 0 
\end{array}\right)\left(\begin{array}{c}
\psi_{b,x}\\
\psi_{b,x-1}
\end{array}\right)=\left(\begin{array}{c}
\psi_{b,x+1}\\
\psi_{b,x}
\end{array}\right),\nonumber\\
F_{k_y,a}\left(\begin{array}{c}
\psi_{a,x}\\
\psi_{a,x-1}
\end{array}\right)=\left(\begin{array}{cc}
\frac{(t_3+t'_3e^{ik_y})(t_4+t'_4e^{-ik_x})-t_1t_2-t'_1t'_2
}{t'_1t_2}
 & \frac{-t_1t'_2}{t'_1t_2} \\
1 & 0 
\end{array}\right)\left(\begin{array}{c}
\psi_{a,x}\\
\psi_{a,x-1}
\end{array}\right)=\left(\begin{array}{c}
\psi_{a,x+1}\\
\psi_{a,x}
\end{array}\right).\nonumber\\
\end{eqnarray} 

With these preparation, we can now analyze the existence of 1D edge states. For a given transfer matrix $F$ with $F(\psi_i,\psi_{i-1})^T=(\psi_{i+1},\psi_i)^T$, assuming $F(u,v)^T=\epsilon(u,v)^T$, we can always rewrite the vector $(\psi_1,\psi_{0})^T$ as a linear combination of the two eigenvector of $F$, i.e. $(\psi_1,\psi_{0})^T=s_1(u_1,v_1)^T+s_2(u_2,v_2)^T$, with $s_1$ and $s_2$ the superposition coefficients. Thus we shall have
\begin{eqnarray}
\left(\begin{array}{c}
\psi_{i+1}\\
\psi_{i}
\end{array}\right)=
s_1\epsilon_1^i\left(\begin{array}{c}
u_1\\
v_1
\end{array}\right)+
s_2\epsilon_2^i\left(\begin{array}{c}
u_2\\
v_2
\end{array}\right).
\end{eqnarray}
Therefore, both of the two eigenvalue $\epsilon_{1,2}$ need to be smaller than unity to have $\psi_i$ localizing around $i=1$, and exponentially decaying with increasing $i$. On the other hand, when both the eigenvalues satisfy $|\epsilon_{1,2}|>1$, one can obtain a solution of $\psi_i$ localizing at the other edge.

The eigen-equation of the transfer matrices read:
\begin{eqnarray}
f_{k_x,b}(\epsilon_{F_b})=t_3t'_4\epsilon_{F_b}^2-[(t_1+t'_1e^{-ik_x})(t_2+t'_2e^{ik_x})-t_3t_4-t'_3t'_4]\epsilon_{F_b}+t'_3t_4=0,\nonumber\\
f_{k_x,a}(\epsilon_{F_a})=t'_3t_4\epsilon_{F_a}^2-[(t_1+t'_1e^{ik_x})(t_2+t'_2e^{-ik_x})-t_3t_4-t'_3t'_4]\epsilon_{F_a}+t_3t'_4=0,\nonumber\\
f_{k_y,b}(\epsilon_{G_b})=t_1t'_2\epsilon_{G_b}^2-[(t_3+t'_3e^{-ik_y})(t_4+t'_4e^{ik_y})-t_1t_2-t'_1t'_2]\epsilon_{G_b}+t'_1t_2=0,\nonumber\\
f_{k_y,a}(\epsilon_{G_a})=t'_1t_2\epsilon_{G_a}^2-[(t_3+t'_3e^{ik_y})(t_4+t'_4e^{-ik_y})-t_1t_2-t'_1t'_2]\epsilon_{F_a}+t_1t'_2=0.
\end{eqnarray}
In our system, we expect to observe only the corner states at zero-energy, which requires that the two eigenvalues of each of these four equations cannot be both larger (or smaller) than unity for any given $k_x$ or $k_y$. However, this is not possible for the case with two diagonal corner states. Consider the cases with $k_{x,y}=0$ and $\pi$, where the coefficients of these equations are all real. Then for each equation, if it has one eigenvalue larger then unity and the other smaller then unity, its corresponding polynomial $f(z)$ must have opposite signs at $z=\pm1$, i.e. $f(1)f(-1)<0$. At $k_{x,y}=0$ and $\pi$, we have:
\begin{eqnarray}
f_{k_x=0,b}(1)&=&f_{k_x=0,a}(1)=(t_3+t'_3)(t_4+t'_4)-(t_1+t'_1)(t_2+t'_2),\nonumber\\
f_{k_x=0,b}(-1)&=&f_{k_x=0,a}(-1)=(t_1+t'_1)(t_2+t'_2)-(t_3-t'_3)(t_4-t'_4),\nonumber\\
\nonumber\\
f_{k_x=\pi,b}(1)&=&f_{k_x=\pi,a}(1)=(t_3+t'_3)(t_4+t'_4)-(t_1-t'_1)(t_2-t'_2),\nonumber\\
f_{k_x=\pi,b}(-1)&=&f_{k_x=\pi,a}(-1)=(t_1-t'_1)(t_2-t'_2)-(t_3-t'_3)(t_4-t'_4),\nonumber\\
\nonumber\\
f_{k_y=0,b}(1)&=&f_{k_y=0,a}(1)=(t_1+t'_1)(t_2+t'_2)-(t_3+t'_3)(t_4+t'_4),\nonumber\\
f_{k_y=0,b}(-1)&=&f_{k_y=0,a}(-1)=(t_3+t'_3)(t_4+t'_4)-(t_1-t'_1)(t_2-t'_2),\nonumber\\
\nonumber\\
f_{k_y=\pi,b}(1)&=&f_{k_y=\pi,a}(1)=(t_1+t'_1)(t_2+t'_2)-(t_3-t'_3)(t_4-t'_4),\nonumber\\
f_{k_y=\pi,b}(-1)&=&f_{k_y=\pi,a}(-1)=(t_3-t'_3)(t_4-t'_4)-(t_1-t'_1)(t_2-t'_2).\nonumber
\end{eqnarray}
Requiring each pair of them to take opposite signs, we have
\begin{eqnarray}
(t_3+\alpha t'_3)(t_4+\alpha t'_4)>(t_1+\beta t'_1)(t_2+\beta t'_2)\label{con1}
\end{eqnarray}
or
\begin{eqnarray}
(t_3+\alpha t'_3)(t_4+\alpha t'_4)<(t_1+\beta t'_1)(t_2+\beta t'_2)\label{con2}
\end{eqnarray}
with $\alpha$ and $\beta$ take $\pm1$.

In the case with two diagonal corner states, based on previous discussion, we need to have two winding numbers $\nu_i=1$, one has $i=1$ or $2$ and one has $i=3$ or $4$, and the other two winding numbers are $t_j=0$. In other words, $|t_i|<|t'_i|$ and $|t_j|>|t'_j|$. Here we take $i=2$ and $4$, and $j=1$ and $3$ as an example, which leads to
\begin{eqnarray}
{\rm sgn}[t_1+ t'_1]&=&{\rm sgn}[t_1- t'_1]={\rm sgn}[t_1],\nonumber\\
{\rm sgn}[t_3+ t'_3]&=&{\rm sgn}[t_3- t'_3]={\rm sgn}[t_3],\nonumber\\
{\rm sgn}[t_2+ t'_2]&=&-{\rm sgn}[t_2- t'_2]={\rm sgn}[t'_2],\nonumber\\
{\rm sgn}[t_4+ t'_4]&=&-{\rm sgn}[t_4- t'_4]={\rm sgn}[t'_4].\nonumber
\end{eqnarray}
Therefore, $(t_3+ t'_3)(t_4+ t'_4)$ and $(t_3- t'_3)(t_4- t'_4)$ shall have different signs, and so do $(t_1+ t'_1)(t_2+ t'_2)$ and $(t_1- t'_1)(t_2- t'_2)$. Hence we can see that either of the Eqs. (\ref{con1}) and (\ref{con2}) cannot be satisfied for all the choices of $\alpha$ and $\beta$. That is to say, 
the transfer matrices must have two eigenvalues both larger (or smaller) then unity, i.e. the system must have 1D edge states, in at least one of the following four cases:
\begin{eqnarray}
{\rm OBC~along~}x,~k_y=0;\nonumber\\
{\rm OBC~along~}x,~k_y=\pi;\nonumber\\
{\rm OBC~along~}y,~k_x=0;\nonumber\\
{\rm OBC~along~}y,~k_x=\pi.\nonumber
\end{eqnarray}

\section{phase boundary}\label{App_phaseboundary}
Due to the time reversal symmetry $H^*(\bm k)=H(-{\bm k})$, the system in the NPSM phase must have gapless points in pairs of ${\bm k}_\pm$, with ${\bm k}_+=-{\bm k}_-$. 
such a pair can only emerge or annihilate in pair at the four high-symmetric points with $k_{x(y)}=0$ or $\pi$,. The phase boundary in Fig. \ref{fig3} can thus be determined accordingly.
The Hamiltonian in momentum space reads
\begin{eqnarray}
H_{k_x-k_y} = \hat{\psi}^\dagger_{\bm k} h({\bm k}) \hat{\psi}_{\bm k},
\end{eqnarray}
where $\hat{\psi}_{\bm k}=(\hat{a}_{\uparrow,{\bm k}} ,\hat{a}_{\downarrow,{\bm k}} , \hat{b}_{\uparrow,{\bm k}} ,\hat{b}_{\downarrow,{\bm k}} ,)^T $ and
\begin{eqnarray}
h(k) &=& \left(\begin{array}{cccc}
0 & 0 & t_1+t'_1e^{-ik_x} & t_3+t'_3e^{-ik_y} \\
0 & 0 & t_4+t'_4e^{ik_y} &  t_2+t'_2e^{ik_x} \\
t_1+t'_1e^{ik_x} & t_4+t'_4e^{-ik_y} & 0 & 0 \\
t_3+t'_3e^{ik_y} & t_2+t'_2e^{-ik_x} & 0 & 0 
\end{array}\right)\nonumber\\
&=&U_{1,+}\tau_1\sigma_0+U_{1,-}\tau_1\sigma_3+V_{1,-}\tau_2\sigma_0+V_{1,+}\tau_2\sigma_3+U_{2,+}\tau_1\sigma_1+U_{2,-}\tau_2\sigma_2+V_{2,+}\tau_1\sigma_2+V_{2,-}\tau_2\sigma_1,\label{kxky}
\end{eqnarray}
$\tau_i$ and $\sigma_i$ the Pauli matrices (or identity matrices if $i=0$) in $(a,b)$ and $(\uparrow, \downarrow)$ subspaces respectively, and 
\begin{eqnarray}
&&U_{1,+}=\frac{(t_1+t'_1\cos{k_x})+(t_2+t'_2\cos{k_x})}{2},~~
U_{1,-}=\frac{(t_1+t'_1\cos{k_x})-(t_2+t'_2\cos{k_x})}{2},\nonumber\\
&&V_{1,+}=\frac{t'_1\sin{k_x}+t'_2\sin{k_x}}{2},~~
V_{1,-}=\frac{t'_1\sin{k_x}-t'_2\sin{k_x}}{2},\nonumber\\
&&U_{2,+}=\frac{(t_3+t'_3\cos{k_y})+(t_4+t'_4\cos{k_y})}{2},~~
U_{2,-}=\frac{-(t_3+t'_3\cos{k_y})+(t_4+t'_4\cos{k_y})}{2},\nonumber\\
&&V_{2,+}=\frac{t'_3\sin{k_y}+t'_4\sin{k_y}}{2},~~
V_{2,-}=\frac{t'_3\sin{k_y}-t'_4\sin{k_y}}{2}.
\end{eqnarray}
The eigen-erengies of this Hamiltonian are given by
\begin{eqnarray}
E=\pm\sqrt{\sum_{\alpha=(1,2),\beta=(+,-)}(U^2_{\alpha,\beta}+V^2_{\alpha,\beta})\pm
2\sqrt{\begin{array}{c}
(U_{1,+}U_{1,-}+V_{1,+}V_{1,-})^2+(U_{1,+}V_{2,+}+V_{1,-}U_{2,-})^2 \\
+(U_{1,-}U_{2,-}-V_{1,+}V_{2,+})^2+(U_{1,+}U_{2,+}+V_{1,-}V_{2,-})^2 \\
+(U_{1,-}V_{2,-}-V_{1,+}U_{2,+})^2+(U_{2,+}U_{2,-}-V_{2,+}V_{2,-})^2\label{energy}
\end{array}}}.
\end{eqnarray}
Among the four high-symmetric points, the one with $(k_x,k_y)=(0,0)$ is always gapped, as Eq. (\ref{energy}) at this point always yields $E(0,0)=\pm2\sqrt{2}$. For the rest three points, requiring the energy dispersion $E(k_x,k_y)=0$, we shall obtain
\begin{eqnarray}
\delta_3\delta_4=-1~{\rm for}~ k_x=0,k_y=\pi;\nonumber\\
\delta_1\delta_2=-1~{\rm for}~ k_x=\pi,k_y=0;\nonumber\\
\delta_1\delta_2+\delta_3\delta_4=0~{\rm for}~ k_x=\pi,k_y=\pi,
\end{eqnarray}
which gives us the phase boundaries between semimetallic and insulating phases of the system, e.g. the three solid lines in Fig. \ref{fig3}(a).

\twocolumngrid
\bibliographystyle{apsrev4-1}

\begin{thebibliography}{99}
\bibitem{footnote}
The two ``2D WTI" regimes in the phase diagram preserve 1D topologies along different directions. In one regime the system only have zero-energy edge states when OBC is taken along $y$ direction [as in Fig. \ref{fig3}(b)], and in the other regime zero-energy edge states exist only when OBC is taken along $x$ direction (no shown).
\end{thebibliography}

\end{document}